\documentclass[aps,prd,10pt,twocolumn,nofootinbib,floatfix]{revtex4}
\usepackage[title]{appendix}
\usepackage{epsfig}
\usepackage{bm}
\usepackage{latexsym}
\usepackage{natbib}
\usepackage{url}
\usepackage{dcolumn}
\usepackage{color}
\usepackage{amsfonts,amssymb,amsmath}
\usepackage{graphicx,epsfig}
\usepackage{psfrag}
\usepackage{subfigure}
\usepackage{hyperref}
\hypersetup{colorlinks=true}
\usepackage{mathtools}
\usepackage{enumitem}
\usepackage{float}
\usepackage{mathrsfs}
\usepackage{xcolor}
\hypersetup{colorlinks,  citecolor=red,  linkcolor=red,  urlcolor=blue}
\usepackage{comment}

\usepackage{mathtools}
\usepackage{gensymb}
\begin{document}

\title{KM3-230213A and IceCube Neutrino Events from Metastable Dark Matter of Primordial Black Hole Origin }

\author{
Prabhav Singh\footnote{24RPH003@spt.pdpu.ac.in},
Mansi Dhuria\footnote{Mansi.dhuria@sot.pdpu.ac.in},
Nathanael Varghese Job\footnote{Nathanael.jmscp23@spt.pdpu.ac.in}
}

\affiliation{
Department of Physics, School of Energy Technology, Pandit Deendayal Energy University (PDEU), Gandhinagar-382426, Gujarat, India
}

\begin{abstract}
\noindent 

We investigate a scenario in which the recently observed ultra-high-energy neutrino event KM3-230213A, with a median energy of approximately 220~PeV, as well as the high-energy neutrinos detected by IceCube Observatory, originate from the decay of superheavy dark matter (DM) particles produced through primordial black hole (PBH) evaporation. To establish this connection, we derive constraints on the PBH abundance parameter $\beta$ as a function of the initial PBH mass $M_{\mathrm{BH_0}}$ and DM mass $m_{\mathrm{DM}}$, by considering the bound from the observed relic DM abundance. Using these constraints, we compute the resulting neutrino flux and show that DM masses in the PeV-EeV range can yield neutrinos of comparable energies, capable of accounting for both the KM3-230213A and IceCube events while remaining consistent with the relic abundance constraint. Interestingly, the scenario remains viable over a broad region of parameter space while satisfying existing cosmological and astrophysical bounds. Overall, our results demonstrate that PBH evaporation followed by DM decay provides a consistent and natural explanation for the observed ultra-high-energy neutrino events in the absence of accompanying multimessenger signatures.


\end{abstract}

\maketitle
\section{Introduction}

The study of multi-messenger signatures has always been important for understanding both galactic and extragalactic environments. Among these, neutrino signatures are particularly significant, as neutrinos can be observed even after traveling cosmological distances owing to their extremely weak interactions with Standard Model (SM) particles. This enables them to propagate along straight paths without appreciable absorption or scattering. In contrast, electromagnetic radiation at PeV energies and above is severely attenuated due to interactions of high-energy photons with the Extragalactic Background Light (EBL) and the Cosmic Microwave Background (CMB)~\cite{IceCube:2018dnn,IceCube-Gen2:2020qha,adriani2025potential}. Consequently, the universe becomes opaque to photons at such extreme energies, making neutrino observations a crucial tool for investigating high-energy astrophysical sources.

The KM3NeT observatory (Cubic Kilometre Neutrino Telescope) has recently reported the detection of a very high-energy neutrino event, designated KM3-230213A. KM3NeT observatory comprises of two detectors placed in the Mediterranean Sea, known as ORCA and ARCA. The ARCA telescope, located near Italy, is designed to detect high-energy cosmic neutrinos while ORCA telescope is focused for atmospheric neutrinos. The ARCA telescope observed a muon with an estimated energy of $120^{+110}_{-60}$ PeV produced through the interaction of an even higher-energy neutrino with the surrounding water near the detector. The median neutrino energy corresponding to this event is estimated to be around 220 PeV, with a $90\%$ confidence interval spanning 72 PeV to 2.6 EeV~\cite{km3net2025observation}. This marks the first-ever detection of a neutrino at such extreme energies, of order ($\mathcal{O}(100)$ PeV) ever observed. In comparison, the IceCube observatory had previously detected neutrinos at energies of order $\mathcal{O}(10)$ PeV, making the KM3-230213A event nearly an order of magnitude higher. The reconstructed equatorial coordinates of KM3-230213A are  RA = 94.3$\degree$, dec. = $-7.8\degree$ \cite{km3net2025observation}, placing its origin far from the Galactic Center and offset by approximately $\sim 11\degree$ from the galactic plane~\cite{KM3NeT:galactic_origin}. 
The detection technique in both the KM3NeT and IceCube experiments is to make the incoming neutrinos interact with the water body in the vicinity of several photomultipliers. The water of the Mediterranean Sea for KM3NeT and the glacier ice for the case of IceCube acts as a water body. The interaction of neutrinos with water/ice produces secondary charged particles that emit Cherenkov radiation \cite{km3_vs_Icecube}. The Cherenkov photons are recorded by Digital Optical Modules (DOMs), which converts them into photoelectrons \cite{IceCubeCollaborationSS:SearchforEHEneutrino}.


Several potential sources of such high-energy neutrinos have been explored, including supernova remnants (SNRs). However, even with efficient acceleration mechanisms in SNRs, cosmic-ray protons can typically attain energies of only a few tens of PeV. Another possible production channel involves cosmic-ray interactions, such as proton–proton ($pp$) and proton–photon ($p\gamma$) collisions. These processes can generate high-energy neutrinos only if the primary protons themselves are accelerated to even higher energies, reaching up to the EeV scale~\cite{KM3NeT:galactic_origin}. Since neither our Milky Way nor nearby galaxies are energetic enough to accelerate protons to such extreme energies, the KM3-230213A event is unlikely to have a galactic origin \cite{Mondol:2025uuw}. Although no definitive source has yet been identified for this high-energy event, several possible origins have been explored. One such possibility is cosmogenic sources, where ultra-high-energy cosmic rays interact with background photons and produce cosmogenic neutrinos\cite{Scrutizing_cosmogenicsourcekm3}. Another likely source is Active Galactic Nuclei (AGN), particularly Blazars. Blazars are a subclass of AGN with relativistic jets aligned close to our line of sight. These jets act as natural cosmic accelerators, boosting the energies of neutrinos and radiation through relativistic Doppler effects~\cite{IceCube:2018dnn,Mondol:2025uuw}. Furthermore, protons ejected from the jets can interact with background photons, producing pions that subsequently decay into high-energy neutrinos through (\textit{p$\gamma$} reaction)  \cite{petrucci2024investigating}.
Another important aspect to address is the discrepancy between the high-energy neutrino flux observed in the KM3–230213A event and that measured by IceCube. While KM3-230213A has reported neutrinos with energies of order 100 PeV, the IceCube Neutrino Observatory has not observed any events beyond a few PeV. This discrepancy is particularly striking given IceCube’s larger effective area and its longer operational exposure, making the absence of comparable detections difficult to reconcile. Moreover, no correlated signals in cosmic rays or gamma rays have been reported by the Pierre Auger Observatory or other facilities. These null observations indicate that the source of KM3-230213A must either involve mechanisms that suppress the production of high-energy photons or be located at cosmological distances, where photons and charged particles are efficiently absorbed or scattered by the intervening extragalactic medium~\cite{Airoldi:2025opo,Fang:2025gammaraycascade}.

A wide range of Beyond Standard Model (BSM) scenarios have been explored to account for the origin of the KM3-230213A event, including DM decay to SM particles~\cite{VectorDMDecayKM3,SHDM_constaintKM3,DebashishBorah,DMdecay_GW_signals_KM3,Kohri:DMdecaykm3,Barman:2025_1,Barman:2025_2,khan2025decayingvectordarkmatter}, attenuation of the neutrino flux through DM-neutrino interactions~\cite{DMnuscattering,DMnuscattering2}, interpretations of the detected muon as a DM signature rather than a neutrino~\cite{KM3notfromneutrino}, production from accelerated right-handed (sterile) neutrinos~\cite{sterile_neutrino_source}, cosmogenic sources~\cite{Scrutizing_cosmogenicsourcekm3,CN_sourcekm3Zhang}, binary black hole mergers associated with electroweak vacuum turbulence ~\cite{BH_merger_KM3}, proton acceleration in blazars~\cite{Accretion_Flare_Blazar_km3,Mondol:2025uuw}, and evaporation of Primordial Black Holes (PBHs)~\cite{pNGB_DM_PBH_KM3,Klipfel:2025jql,memory_burden_KM3,Airoldi:2025opo} etc.

Among these possibilities, PBHs present a particularly intriguing source of high-energy neutrinos. Beyond their potential role in addressing outstanding questions in cosmology, such as the origin of DM~\cite{Khlopov_4,Khlopov_5,Khlopov_6},  microlensing anomalies near the galactic bulge~\cite{PhysRevD.OGLE_microlensing}, and the apparent absence of ultra-faint dwarf galaxies \cite{PBH_book,SevenhintsPBH}, they also provide a natural mechanism for producing energetic particles through Hawking evaporation. Constraints from Big Bang Nucleosynthesis (BBN)  require PBHs in the mass range 
 $10^9{\rm g}\leq m_{\rm PBH} \leq 10^{14}{\rm g}$
to have evaporated before nucleosynthesis~\cite{cheek2022primordial}, while inflationary considerations impose a lower bound of $m_{\rm PBH}\gtrsim 0.1$g~\cite{cheek2022primordial,carr2021constraints,keith2020constraints,carr2010_constraint_PBH}. PBHs with  $m_{\rm PBH} \leq 10^{14}{\rm g}$ are expected to have evaporated by the present epoch, releasing their most energetic particles near the end of their lifetime, when the evaporation rate rises sharply~\cite{hawking1974blackexplosion}.  Since Hawking radiation is universal, all kinematically allowed species, both SM and beyond, can be produced, irrespective of their interactions~\cite{carr1974black,hawking1974blackexplosion,Hawking_mnras_gravitational_collapse_PBH,carr_mass_spectrum,khlopov_2}. The resulting relics may leave diverse cosmological imprints. Massless relics contribute to the radiation background, whereas massive relics may later become non-relativistic and act as dark matter~\cite{warm_Hawking_relics_shallue}. 
 
 Of particular importance for the present work is the emission of neutrinos from PBH evaporation, which can arise through two distinct channels. First, PBHs can directly radiate neutrinos thermally via Hawking radiation. Second, PBHs can emit long-lived, superheavy DM that subsequently decays into neutrinos at later times. Various ideas have been explored regarding PBHs as viable sources for the KM3–230213A event. These include models involving neutrino emission from pseudo-Nambu-Goldstone (pNGB) DM particles evaporated through PBH Hawking radiation \cite{pNGB_DM_PBH_KM3}, high energy neutrinos from the decay of superheavy sterile neutrino produced from Hawking radiation by employing type-1 seesaw mechanism~\cite{choi2025sterile}, direct neutrino emission via Hawking radiation from memory–burdened PBHs constituting a fraction of DM~\cite{Klipfel:2025jql}, and predictions for future neutrino events expected in the KM3NeT detector \cite{memory_burden_KM3}. Some studies have also disfavored the PBH origin hypothesis by suggesting that, to account for the observed event, the PBH would need to be located within the solar system-an unlikely scenario given the absence of corresponding multimessenger signals \cite{Airoldi:2025opo}.



In this study, we focus on the indirect mechanism in which neutrinos are produced from the decay of DM particles produced through PBH evaporation. Such decays naturally yield neutrinos with energies higher than those of thermally produced ones, making this scenario a compelling framework to explain the observed ultra-high-energy neutrino events. Though the neutrino emission from PBHs via Hawking radiation has been already discussed in the literature in the context of KM3-230213A event, the interplay between this process and the relic abundance of DM has not been systematically incorporated. Since both the neutrino flux and the relic abundance of DM are sensitive to PBH parameters such as the initial mass ${M_{\rm BH}}_0$ and the PBH fractional energy density $\beta$, it is essential to verify the consistency of the parameter space allowed by observational constraints on both quantities. Assuming that superheavy DM is produced non–thermally through PBH evaporation, we employ the measured relic abundance of DM from the Planck satellite to constrain the PBH abundance and the DM yield. Using these constraints, we compute the neutrino flux arising from the decay of the superheavy DM and find that neutrinos in the PeV–EeV energy range can be produced for appropriate DM masses across a broad range of PBH parameters. Finally, we demonstrate that this indirect emission mechanism can successfully account for both the KM3–230213A and IceCube high-energy neutrino events within a consistent region of the parameter space.

The paper is organised as follows. In Sec. \ref{PBH}, we review the formation and evaporation of PBHs and discuss their evolution through Hawking radiation. In Sec. \ref{Relic abundance of PBH and DM}, we constrain the PBH abundance  $\beta$ as a function of PBH and DM masses through consistency with the observed relic abundance of DM. In Sec. \ref{Neutrino flux from SHDM}, we compute the neutrino flux from the decay of superheavy DM produced via Hawking radiation from PBHs. In Sec. \ref{KM3 Interpretation}, we present our main results and argue that the decay of superheavy DM originating from PBHs offers a viable explanation for the recently observed KM3NeT neutrino event, while remaining consistent with the observed relic abundance of DM. In Sec. \ref{Conclusions}, we provide a summary of the main results and outline the broader implications of this work.

\section{Primordial Black Hole}
\label{PBH}
\subsection{Formation}
The inhomogeneities in the small scale in the early universe, if large enough, would lead to the formation of primordial black holes. The mass of these primordial black holes would relate to the time after the Big Bang as~\cite{carr2021constraints}:
\begin{equation}
\label{mass_time_eq}
M \sim \frac{c^3t}{G}\sim 10^{15} \left(\frac{t}{10^{-23}s}\right)g.    
\end{equation}
This means that at the Planck time ($10^{-43}$s), a black hole of mass $10^{-5}$g could be formed. A black hole formed one second after the Big Bang would have a mass of about $10^5{\rm M_{\odot}}$. 

If PBHs are formed due to gravitational collapse of primordial perturbations, the mass of the PBH at the formation time, ${\rm M_{BH_{0}}}$, is expected to be of the order of ${\rm M_H}$ where ${\rm M_H}$ is the mass contained within the Hubble volume at the time of black hole formation. The relation can be written as ${\rm M_{BH_0}}=\gamma {\rm M_H}$. 
If we assume the black hole behaves as a Schwarzschild black hole, its mass in relation to the volume can be written as M = density $\times$ volume. For a black hole in the radiation era at plasma temperature $T_{F}$:
\begin{equation}
\label{mass_density_eq}
    M_{\rm BH_0}(T_F) =\frac{4}{3}\pi \gamma \frac{\rho_{\rm rad}(\rm T_F)}{H_F^3(\rm T_F)},
\end{equation} where $\gamma \approx 0.2$ is a dimensional parameter denoting gravitational collapse, $H_F=1/(2t_F)$ is the Hubble scale with $t_{F}$ being the formation time, and  $\rho_{\rm rad}$ is the energy density of BH in radiation era, given by \cite{OGcontourpaper,NonCDMpaper}:
\begin{equation}
    \label{rho_rad}
    \rho_{\rm rad}(T) = \frac{\pi^2}{30}g_*(T)T^4.
\end{equation}
Using the Friedmann equation $H_F^2 = \frac{8\pi G\rho_{\rm rad}}{3}$, we get:
\begin{equation}
\label{time_mbh0_eq}
    H_F^2 = \frac{8\pi \rho_{\rm rad}}{3M_p^2} = \frac{\pi}{3}\sqrt{\frac{8\pi g_*(T)}{10}}\frac{T^2}{M_p} 
\end{equation}
By comparing this with $H_F=1/(2t_F)$, we obtain the formation time of the primordial black hole as:
\begin{equation}
\label{time t_F}
t_F = \frac{M_{\rm BH_0}}{\gamma M_p^2},
\end{equation}
where $M_{BH_0}$ is the black hole mass at formation time and $M_p=\sqrt{1/G} = 1.22\times10^{19}$ GeV is the Planck mass.
Inflation sets the lower bound constraint for the initial black hole mass~\cite{NonCDMpaper}:
\begin{equation}
    \label{inflation_constraint_mbh0}
    M_{\rm BH_0} \gtrsim 10^4 M_p.
\end{equation}

\subsection{Evaporation}
According to Hawking radiation, a non-rotating black hole with zero charge would emit particles of $i^{\rm th}$ species explained by the Hawking radiation rate~\cite{NonCDMpaper,HEandUHEchina,OGcontourpaper}:
\begin{equation}
    \label{Hawking_rad_rate}
    \frac{d^2 N_i}{dtdE}\approx \frac{g_i}{2\pi}\cdot \frac{\gamma_{\rm grey}}{exp(E/T_{\rm BH})\pm 1},
\end{equation}
where $N_{i}$ is the number of particles $i$ emitted at an average rate per energy interval, $g_{i}$ is the multiplicity of particle $i$, and $\gamma_{\text{grey}}$ is the greybody factor, which depends on the particle energy, the mass of the black hole, and the particle’s properties.The temperature of the black hole depends inversely on its mass and is given by
\begin{equation}
    \label{temp_mbh0_eq}
    T_{\rm BH} = \frac{M_p^2}{8\pi M_{\rm BH}}.
\end{equation}
From eq.~\eqref{mass_time_eq}, black holes formed later in the universe would have larger masses. Such black holes would not have completely evaporated by the present time. Since we are interested in black holes that evaporated long ago, we restrict our attention to lighter-mass black holes.

Eqs.~\eqref{Hawking_rad_rate} and \eqref{temp_mbh0_eq} together imply that the mass of a black hole decreases with time at the rate~\cite{NonCDMpaper}:
\begin{equation}
    \label{mass_evolution}
    \frac{dM_{\rm BH}}{dt} = -\sum_j \int_0^\infty E \frac{dN_i}{dtdE}dE = -e_T\frac{M_p^4}{M_{\rm BH}^2},
\end{equation}
with $
    e_T = \frac{27}{4}\frac{g_{*BH}}{30720\pi}$.
Solving this, one gets:
\begin{equation}
    \label{mass evolution 2}
    M_{\rm BH}(t) = M_{\rm BH_0}\left(1-\frac{(t-t_F)}{\tau}\right)^{1/3},    
\end{equation}
where $M_{BH_0}$ is the black hole mass at formation time $t_{F}$, and $\tau$ is the lifetime of the black hole, expressed as
\begin{equation}
    \label{PBH_lifetime}
    \tau = \frac{1}{3e_T}\frac{M_{\rm BH_0}^3}{M_p^4}
\end{equation}
The total number of particles of species $i$ emitted during PBH evaporation is obtained by integrating Eq.~\eqref{Hawking_rad_rate} over the PBH lifetime $\tau$ and the total energy $E$:  
\begin{equation}
    N_i = \int_0^{\tau} \! dt \int_0^{\infty} \! \frac{d^2 N_i}{dE\, dt} \, dE \, .
\end{equation}
Carrying out the integration yields~\cite{HEandUHEchina}  
\begin{align}
    \label{N_DM1}
    N_i = C_n \frac{120 \zeta(3)}{\pi^3}\frac{g_i}{g_*(T_F)}\left(\frac{M_{BH_0}}{m_p}\right)^2,~ m_i<T_{F}\\ 
    \label{N_DM2}
  \hskip -0.6in {= C_n \frac{15 \zeta(3)}{8\pi^5}\frac{g_i}{g_*(T_F)}\left(\frac{m_p}{m_i}\right)^2,~ m_i>T_{F}.}
\end{align}
The temperature of the plasma surrounding the black hole at the formation time is given by~\cite{NonCDMpaper}:
\begin{equation}
    \label{plasma_temp}
    T_{\rm plasma}(t_F) = \left(\sqrt{\frac{45}{16\pi^3g_*(t_F)}}\frac{\gamma M_p^3}{M_{\rm BH_0}}\right)^{1/2}
\end{equation}
It is important to note that this plasma temperature is higher than the initial black hole temperature $T_F$ given in eq.~\eqref{temp_mbh0_eq}. This implies that primordial black holes do not start evaporating immediately; instead, they begin evaporation only after the plasma temperature cools down or once the PBHs come to dominate the energy density of the Universe~\cite{NonCDMpaper}.

\section{Relic abundance of PBH and DM}
\label{Relic abundance of PBH and DM}
Primordial black holes, if formed in the early universe, can contribute significantly to the present-day DM abundance. To quantify this contribution, one typically introduces the initial abundance parameter $\beta$, defined as the ratio of the PBH energy density to the radiation energy density at the time of PBH formation. Since PBHs can evaporate into all particle species, including DM, $\beta$ serves as the key parameter for estimating both the relic abundance of PBHs and the DM produced through their evaporation~\cite{HEandUHEchina}. In the following, we outline and review the calculation steps leading to the determination of the DM relic abundance sourced by PBH evaporation. To begin with, the parameter $\beta$ is expressed as~\cite{HEandUHEchina}
\begin{equation}
    \label{beta_omega}
    \beta \equiv \Omega_{\rm PBH}(t_{\rm F}) = \frac{\rho_{\rm PBH}}{\rho_{\rm rad}}.
\end{equation}
where $t_{\rm F}$ denotes the formation time. Accordingly, the total number density of PBHs at formation time ($t_{\rm F}$) is given by~\cite{NonCDMpaper},
\begin{equation}
    \label{number_dens_PBH}
    n_{\rm PBH}(t_F) = \frac{\beta}{M_{\rm BH_0}}\cdot \rho_{\rm rad}(t_{\rm F}).
\end{equation}
Measurements of the Cosmic Microwave Background (CMB) by the Planck satellite constrain the DM relic abundance to $\Omega_{\rm DM}h^2 = 0.120 \pm 0.001$~\cite{Planck_sat_DM}. This abundance can be expressed in terms of the conserved yield $Y_{\rm DM} \equiv n_{\rm DM}/s$, where $n_{\rm DM}$ is the DM number density and $s$ is the entropy density. In terms of $Y_{\rm DM}$, the DM relic abundance is given by  

\begin{equation}
    \label{relic_DM_1}
    \Omega_{\rm DM}h^2 = \frac{\rho_{\rm {DM,0}}}{\rho_c} = \frac{s_0}{\rho_c}m_{\rm DM}Y_{\rm DM},
\end{equation}
where $s_0$ is the present-day entropy density and $\rho_c$ is the critical density, given by
\begin{align}
    \label{relic_DM_constants}
    s_0=\frac{2891.2}{\rm cm^3},~
    \rho_c=1.0537\times10^{-5}h^2 \frac{\rm GeV}{\rm cm^3}.
\end{align}  
Substituting these values, one finds
\begin{equation}
    \label{relic_ab_DM}
    \Omega_{DM}h^2 = 2.74385 \times 10^8 \frac{m_{\rm DM}}{\rm GeV}Y_{\rm DM}.
\end{equation}

Since $Y_{\rm DM}$ is a conserved quantity and does not depend on the scale factor, it can be calculated as
\begin{equation}
\label{eq:YDM_RD1}
Y_{\rm DM} = \frac{ n_{\rm DM}(T_{\rm F})}{s(T_{\rm F})} = \frac{N_{\rm DM}~  n_{\rm PBH}(T_{\rm F})}{s(T_{\rm F})}.
\end{equation}
where $n_{\rm DM}({\rm T_ F})$ and $n_{\rm PBH}({\rm T_F})$ are the number densities of DM particles and PBHs, respectively, at the time of PBH formation, ${\rm T_F}$ represents the temperature at the time of formation of PBH, and ${\rm N}_{\rm DM}$ represents the total number of DM particles emitted during the evaporation of PBH. 

Depending on the initial abundance, PBHs can dominate over radiation and induce an early matter-dominated era. Accordingly, the expression for the comoving number density depends on whether PBH evaporation takes place during radiation domination or matter domination. 
Since matter and radiation scale differently with temperature, we define the equality temperature $T_{\rm eq}$ through the condition $\rho_{\rm BH}(T_{\rm eq}) = \rho_{\rm rad}(T_{\rm eq})$. This gives
\begin{equation}
    T_{\rm eq} = \beta \, T_{\rm F}
    \left( \frac{g_{\star,s}(T_{\rm F})}{g_{\star,s}(T_{\rm eq})} \right)^{1/3},
    \label{eq:Teq}
\end{equation}
where $T_{\rm F}$ is the PBH formation temperature.  
At temperatures below $T_{\rm eq}$, the expansion of the universe becomes PBH-dominated, since $\rho_{\rm PBH} > \rho_{\rm rad}$. To characterize this transition, one introduces the critical parameter $\beta_c$, defined as
\begin{equation}
    \beta_c \equiv \frac{T_{\rm eq}}{T_{\rm F}} \equiv \frac{T_{\rm ev}}{T_{\rm F}},
    \label{eq:betac}
\end{equation}
where $T_{\rm ev}$ denotes the PBH evaporation temperature. For $\beta < \beta_c$, PBHs evaporate in the radiation-dominated era, whereas for $\beta > \beta_c$, the universe undergoes an early matter-dominated phase driven by PBHs.  To evaluate $\beta_c$, we need to evaluate $T_{\rm ev}$. 
If PBHs evaporate in the radiation-dominated era, the evaporation temperature follows from $H = 1/(2\tau) =  (T_{\rm ev}^{\rm rad})^2/M_p$, yielding
\begin{equation}
    T_{\rm ev}^{\rm rad} \simeq 
    \frac{\sqrt{3}}{4}
    \left( \frac{g_\star(T_F)}{45} \right)^{1/4}
    \left( \frac{m_p^5}{M_{\rm BH_0}^3} \right)^{1/2},
    \label{eq:Tev_rad}
\end{equation}
where $m_{\rm p} = M_{\rm P}/\sqrt{8\pi}$ is the reduced Planck mass, and $g_\star(T_{\rm F})$ is the number of relativistic degrees of freedom at PBH formation.  

For $\beta > \beta_c$, PBHs dominate before evaporation, and equating $H = 2/(3\tau)$ gives
\begin{equation}
    T_{\rm ev}^{\rm mat} \simeq
    \left( \frac{g_\star(T_{\rm F})}{640} \right)^{1/4}
    \left( \frac{m_p^5}{M_{\rm BH_0}^3} \right)^{1/2}.
    \label{eq:Tev_mat}
\end{equation}

Since $T_{\rm ev}^{\rm rad} \equiv T_{\rm ev}^{\rm mat} \equiv T_{\rm ev}$, the critical fraction $\beta_c$ is finally expressed as $\beta_c = \frac{T_{\rm ev}}{T_{\rm F}}$. Based on the value of $\beta_c$, we can evaluate  $Y_{\rm DM}$ in either the radiation or matter-dominated era. 

If PBHs evaporate during radiation domination, the entropy density at formation is
\begin{equation}
    s(T_{\rm F}) = \frac{2 \pi^2}{45} \, g_{\star,s}(T_{\rm F}) \, T^3_{\rm F},
    \label{eq:entropy_density}
\end{equation}
where $g_{\star,s}(T_{\rm F})$ is the effective number of relativistic degrees of freedom in entropy. Substituting $n_{\rm PBH}(T_{\rm F})$ from Eq.~\eqref{number_dens_PBH} into Eq.~\eqref{eq:YDM_RD1} yields
\begin{equation}
    Y^{\rm rad}_{\rm DM} 
    = \beta \, \frac{N_{\rm DM}}{M_{\rm BH}} \, 
    \frac{\rho_{\rm rad}(T_{\rm F})}{s(T_{\rm F})}.
    \label{eq:YDM_RD2}
\end{equation}
Using Eqs.~\eqref{rho_rad} and \eqref{eq:entropy_density}, this becomes
\begin{equation}
    Y^{\rm rad}_{\rm DM} 
    = \frac{3}{4} \, \frac{g_\star(T_{\rm in})}{g_{\star,s}(T_{\rm in})} \, 
    \beta \, N_{\rm DM} \, \frac{T_{\rm F}}{M_{\rm BH_0}} .
    \label{eq:YDM_RD2}
\end{equation}
The formation temperature of the PBH follows from Eqs.~\eqref{mass_density_eq} and \eqref{time_mbh0_eq} as
\begin{equation}
    T_{\rm F} = \sqrt{3\gamma} 
    \left( \frac{160}{g_\star(T_{\rm F})} \right)^{1/4} 
    \left( \frac{M_{\rm P}^3}{M_{\rm BH_0}} \right)^{1/2}.
    \label{eq:Tf}
\end{equation}
Substituting Eq.~\eqref{eq:Tf} into Eq.~\eqref{eq:YDM_RD2} gives the DM yield in the radiation-dominated era~\cite{NonCDMpaper}:
\begin{equation}
    \label{Y_DM_Rad}
    Y_{\rm DM}^{\rm rad} = \sqrt{3\gamma}\left(\frac{405}{8g_*(T_{\rm F})}\right)^{1/4}\beta N_{\rm DM}\left(\frac{m_p}{M_{\rm BH_0}}\right)^{3/2},
\end{equation}
where $N_{\rm DM}$ is the total number of DM particles emitted per PBH.

If PBHs dominate the cosmic energy density, i.e., for $\beta>\beta_c$, the universe undergoes an early matter-dominated epoch driven by PBHs. In this case, the DM yield from evaporation is given by~\cite{HEandUHEchina}
\begin{equation}
    \label{Y_DM_mat}
    Y_{\rm DM}^{\rm mat} = \frac{3}{8}\left(\frac{g_*(T_{ev})}{40}\right)^{1/4}\left(\frac{m_p}{M_{\rm BH_0}}\right)^{5/2}N_{\rm DM},
\end{equation}
where $g_\star(T_{\rm ev})$ denotes the number of relativistic degrees of freedom at evaporation. 

In addition, the scale factor at evaporation can be determined from comoving entropy conservation, $s a^3=\text{const.}$, which implies $s_0 a_0^3 = s_{\rm ev} a_{\rm ev}^3$, with $s_0$ and $a_0$ ($s_{\rm ev}$ and $a_{\rm ev}$) denoting the entropy density and scale factor today (at evaporation). Employing the entropy density expressions in radiation- and matter-dominated eras, corresponding to $T_{\rm ev}^{\rm rad}$ and $T_{\rm ev}^{\rm mat}$, one finds~\cite{NonCDMpaper}:
\begin{eqnarray}
  \label{a_ev_rad}
   &&  a_{\rm ev}^{\rm rad}
    =2.5\times 10^{-31}\left(\frac{M_{\rm BH_0}}{M_{\rm p}}\right)^{3/2}, \nonumber\\
&& 
    a_{\rm ev}^{\rm mat} = 2.2\times 10^{-31}\left(\frac{M_{\rm BH_0}}{M_{\rm p}}\right)^{3/2}.
\end{eqnarray}

\paragraph*{Relic abundance of DM:}
We now evaluate the relic abundance of DM produced through the evaporation of PBHs. Since our focus is on the flux of neutrinos originating from DM emitted during the radiation-dominated era, we calculate the DM relic abundance specifically for this epoch, using the relevant expressions derived in the preceding discussion.  By substituting Eq.~\eqref{N_DM2} into the expression for the DM yield given in Eq.~\eqref{Y_DM_Rad}, and subsequently inserting the result into the relic abundance relation in Eq.~\eqref{relic_ab_DM}, we obtain the expression for the DM relic abundance in the radiation-dominated era as:
\begin{equation}
    \label{omega_RD}
    \Omega_{\rm DM}^{\rm rad}h^2 = 0.12 \left(\frac{\beta}{2.72\times 10^{-24}}\right) \, \left(\frac{400\, \rm PeV}{m_{\rm DM}}\right) \, \left(\frac{10 \, \rm g}{M_{\rm BH_0}}\right)^{3/2}.
\end{equation}

\begin{figure}[htbp]
    \includegraphics[width = \linewidth]{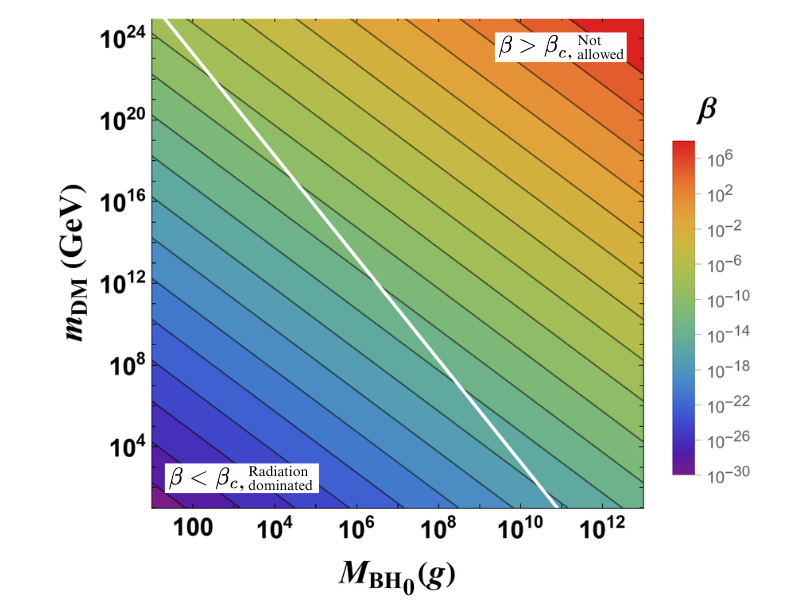}
    \caption{Constraints on the PBH abundance, $\beta$, as a function of the black hole mass $M_{\rm BH_0}$ and DM mass $m_{\rm DM}$, obtained in the radiation-dominated era using the observed relic abundance of DM from the \textit{Planck} satellite, $\Omega_{\rm DM} h^2 = 0.12$. The white line corresponds to the critical abundance $\beta = \beta_c$. The region to the left of the white line represents the allowed parameter space in the $M_{\rm BH_0}$-$m_{\rm DM}$ plane for a given range of $\beta$ during the radiation-dominated era.  The region to the right of the white line is not allowed.}
    \label{contourplot}
\end{figure}
In Fig.~\ref{contourplot}, we present the constraints on the PBH abundance as a function of the initial black hole mass and the corresponding DM mass produced from PBH evaporation in the radiation-dominated era ($\beta < \beta_c$). The results indicate that superheavy DM can be obtained for black hole masses in the range $10~\mathrm{g}$ to $10^{11}~\mathrm{g}$, for a wide range of $\beta$ between $10^{-6}$ and $10^{-30}$.  In the next section, we investigate whether the neutrinos arising from the decay of such superheavy DM can reproduce the observed flux measured by the IceCube and KM3NeT detectors, while remaining consistent with the allowed parameter space.

\section{Neutrino flux from the decay of Superheavy dark matter}
\label{Neutrino flux from SHDM}

In this section, we focus on calculating the neutrino flux produced by the evaporation of PBHs. As PBHs evaporate, they emit high-energy particles, with their temperatures steadily increasing and potentially approaching the Planck scale in the final stages of the process. The resulting particles can be either directly emitted by the PBH or generated through the decay of heavy particles in the early universe. To evaluate the corresponding flux, one must determine the phase-space distribution of the emitted particles. In an expanding universe, this evolution is governed by the Boltzmann equation~\cite{HEandUHEchina},
\begin{equation}
    \label{Boltzmann_eq}
    \left[\frac{\partial}{\partial t}-Hp\frac{\partial}{\partial p}\right]f_i(t,p)\approx \Gamma_{i,\rm prod}-\Gamma_{i,\rm abs}f_i
\end{equation}
where $f_i(t, p)$ is the phase-space distribution function of particle species ``$i$'', 
$H = \dot{a}/a$ is the Hubble parameter, $\Gamma_{i,\text{prod}}$ denotes the production 
rate of particle species $i$ (including both thermal processes and PBH evaporation), 
and $\Gamma_{i,\text{abs}}$ is the absorption rate in the plasma.

High-energy neutrinos can be produced from PBH evaporation through two distinct mechanisms. In the direct mechanism, neutrinos are emitted as part of the Hawking radiation spectrum. In the indirect mechanism, PBHs emit superheavy DM particles, which subsequently decay into neutrinos at a later stage. In the case of direct neutrino production, since these neutrinos are generated very early in the universe, their energies are significantly affected by several processes such as cosmological redshift, thermal wash-out at the high-energy end, Hubble expansion, and interactions with the dense thermal plasma~\cite{HEandUHEchina}. By contrast, in the indirect case,  neutrinos produced from the decay of DM particles are largely unaffected by these effects as they are generated much later in cosmic history. 
This has been shown in Ref.~\cite{HEandUHEchina} that for a wide range of initial masses of a black hole, the flux from direct neutrino production (solid line) gets cut off towards the high-energy end near the neutrino energy of $10^6\,\rm GeV$. Since our interest lies in explaining the origin of the very high-energy neutrino regime in the range 100 PeV - EeV, we therefore focus on the indirect production channel, namely, neutrinos originating from the decay of superheavy dark matter. We require the decay width to satisfy
\begin{equation}
    \label{decaying_cond}
    \Gamma_{iD} \ll H_{ev},
\end{equation}
where $\Gamma_{iD}$ is the decay rate of particle ``i".

Neglecting the absorption term in Eq.~(\ref{Boltzmann_eq}), we can analytically solve the Boltzmann equation as~\cite{HEandUHEchina},
\begin{equation}
    \label{f_Bolt}
    f_i(t,p) = \int_0^a\frac{\Gamma_{\rm prod}(a',p')}{H(a')a'}da'\,\,\,\,, \,p'=pa/a'
\end{equation}
Considering the decay of DM into lighter particles (for instance, neutrinos) through the process $i\rightarrow j+k$, the Boltzmann equation for the distribution function of particle $j$ can be written as\footnote{For a detailed derivation of eq.~\eqref{f_bolt}, see the appendix of \cite{HEandUHEchina}.}
\begin{equation}
    \label{f_bolt}
    f_j(t,p_j) = \frac{m_i\Gamma_{iD}}{p_j^2a_t^2}\int_{t_{ev}}^ta_\tau^2d\tau\int_{p_i^{min}}^\infty\, dp_i\frac{p_i}{E_i}f_i(\tau,p_i),
\end{equation}
where 
\begin{equation}
    p_i^{min}\equiv \left|\frac{m_ia_\tau}{4p_ja_t}-\frac{p_ja_t}{a_\tau}\right|.
\end{equation}
Since we are considering superheavy DM, we focus on the case where the parent distribution $f_i$ is non-relativistic at the time of decay. In this limit, the phase-space distribution of the daughter particle $j$ (in our case, neutrinos) takes the form~\cite{HEandUHEchina},
\begin{align}
\label{f_j}
f_j(t, p_j) &\approx 16 \pi^2 n^{\rm ev}_{i} a_{ev}^3 
\left( \frac{2 p_j a_t}{m_i a_d} \right)^r 
\left( \frac{1}{2 p_j a_t} \right)^3 \notag \\
  &\quad
\exp \left[ -\frac{1}{r} \left( \frac{2 p_j a_t}{m_i a_d} \right)^r \,\right],
\end{align}
with
\begin{equation}
r = 
\begin{cases}
2 \,\, (a_d \in \text{rad.)}\\
\frac{3}{2} \,\,(a_d \in \text{mat.)},
\end{cases}    
\end{equation}
where $n^{\rm ev}_{i}$ and $a_{ev}$ denote the number density and scale factor at the epoch of evaporation, 
while $a_d$ is the scale factor at the time when the particle $i$ decays into $j$, i.e. when $\Gamma_{iD}=H$. 
The quantity $n^{\rm ev}_{i}$ can be obtained from~\eqref{niev}, and the parameter $r$ is an index that depends 
on whether the decay epoch $a_d$ lies in the radiation-dominated or matter-dominated era, thus governing 
the spectral shape of the distribution. We consider a scenario in which superheavy dark matter undergoes late-time decay during the matter-dominated epoch into neutrinos and lighter dark-sector particles. For definiteness, we fix the parameter $r=3/2$ and substitute the corresponding value of the decay scale factor $a_d$ in our numerical calculations. We further assume that this superheavy species constitutes only a fraction of the total dark matter density of the universe.

Further, in our case, the neutrinos are produced through the decay of DM, 
which itself originates from the evaporation of PBHs. Therefore, the
number density of superheavy DM particles emitted from PBH evaporation, $n_i^{\rm ev}$, 
can be calculated from the bounds on the relic abundance of DM, as given in Eq.~(\ref{relic_ab_DM}). We know that the yield of a particle species $i$ is proportional to the ratio of its number 
density to the entropy density, which is written as
\begin{equation}
    Y_i = \frac{n_i^{\rm ev}}{s_i^{\rm ev}}, 
    \label{eq:Yi_def}
\end{equation}
where $n_i^{\rm ev}$ and $s_i^{\rm ev}$ denote the number density and entropy density 
of species $i$ at the PBH evaporation time. The entropy density can be expressed in terms of the temperature as
\begin{equation}
   s_i^{\rm ev} = \frac{2\pi^2}{45}\, g_{*s}(T)\,(T^{{\rm ev}}_i)^3, 
    \label{eq:entropy}
\end{equation}
where $g_{*s}(T)$ denotes the effective number of relativistic degrees of freedom 
contributing to the entropy. The DM yield is constrained by the observed relic abundance, 
$\Omega_{\rm DM} h^{2} \simeq 0.12$. Substituting the DM yield from Eq.~(\ref{relic_ab_DM}) 
into Eq.~\eqref{eq:Yi_def}, we obtain the number density of superheavy DM particles at 
the PBH evaporation time as
\begin{equation}
\label{niev}
    n^{\rm ev}_i = n_{{\rm DM}} = Y_{\rm DM}\frac{2\pi^2}{45}\cdot (T_{\rm ev}^{\rm rad})^3.
\end{equation}

\begin{figure}
    \centering
    \includegraphics[width=\linewidth]{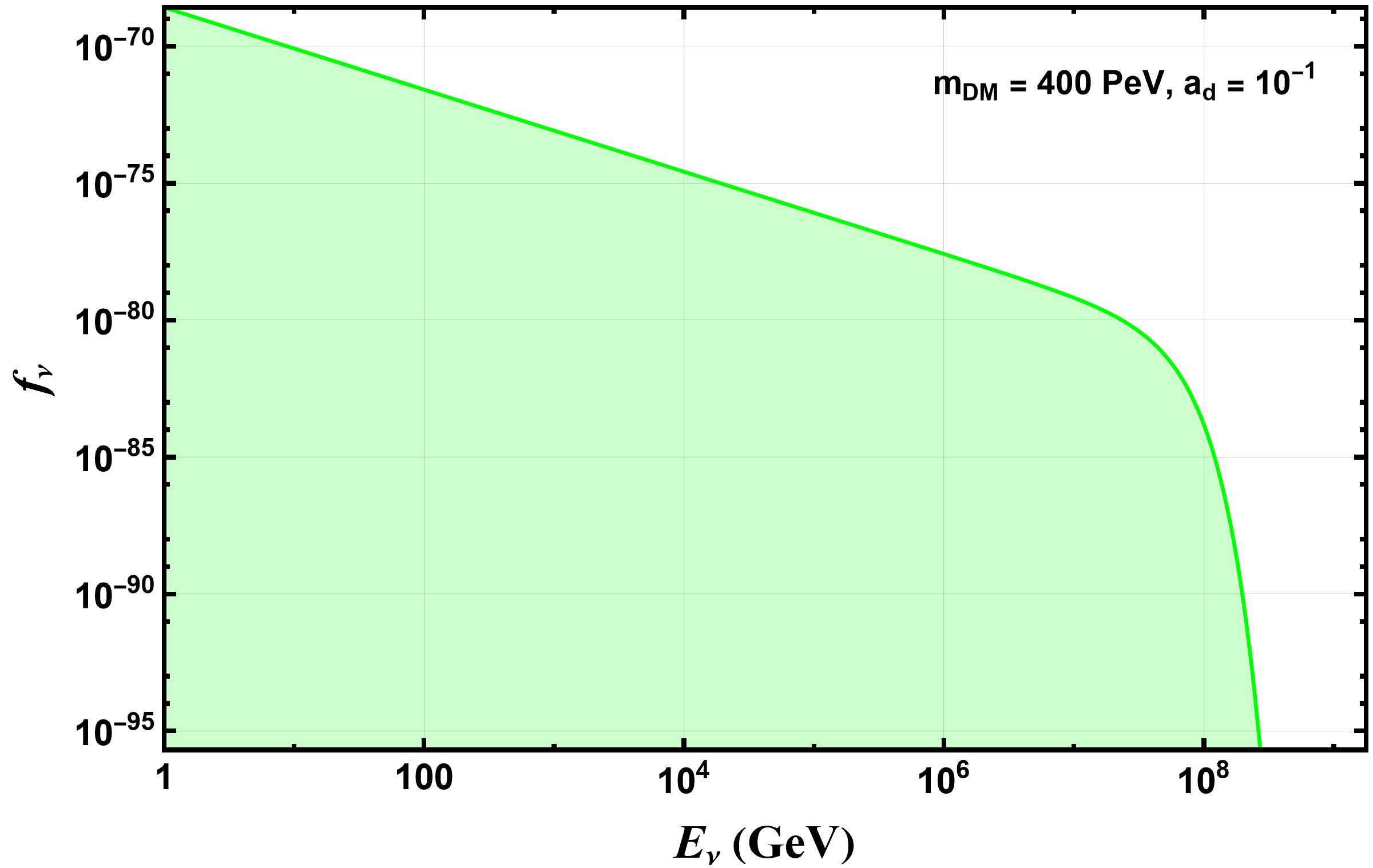}
    \caption{Phase-space distribution function of neutrinos produced from DM decay, shown for fixed values of the primordial black hole mass, DM mass, and the scale factor at which the decay occurs. }
    \label{Neutrino_flux_fnu}
\end{figure}

By substituting the expression for $n_{i}^{\rm ev}$ from Eq.~(\ref{niev}) into Eq.~(\ref{f_j}), we obtain the phase-space distribution of neutrinos produced via PBH evaporation for a given PBH and dark matter mass. Figure~\ref{Neutrino_flux_fnu} shows the neutrino phase-space distribution function resulting from dark matter decay as a function of neutrino energy, for $m_{\rm DM} = 400~{\rm PeV}$, and $a_d = 0.1$. Once the phase-space distribution function of neutrinos, $f_\nu$, is obtained, the differential neutrino flux can be calculated using~\cite{HEandUHEchina},
\begin{equation}
    \label{diff_nu_flux}
    \frac{d\Phi_\nu}{dE_\nu} = \frac{E_\nu^2}{2\pi^2}f_\nu.
\end{equation}
In our case, we obtain:
\begin{equation}
    \frac{d\Phi_\nu}{dE_\nu} = (0.5 \times 10^2) \ \rm f_{DM}\,e^{\frac{-4\sqrt{2}}{3}\left(\frac{E_\nu }{\rm a_d\, m_{DM}}\right)^{3/2}}\left(\frac{E_\nu}{a_d^3\, m_{DM}^5}\right)^{1/2}.
    \label{eq:finflux}
\end{equation}
where $\Phi_\nu$ denotes the number of neutrinos detected per unit time per unit area, $f_{\rm DM}$ is the factor for PBH constituting a fraction of the total DM, and $E_\nu$ represents the neutrino energy. The neutrino flux is expressed in units of $\rm GeV^{-1}cm^{-2}s^{-1}$. 

As we assume the metastable decay of superheavy DM during the matter-dominated era, we equate the Hubble expansion rate in this epoch with the decay rate of the superheavy DM particle. This allows us to express the scale factor at the time of decay in terms of the DM lifetime. The detailed steps of this calculation are provided in Appendix~\ref{appendix:1}. Using those results, we can rewrite the energy flux of the ultra-high-energy neutrino in terms of the dark matter lifetime as follows:
\begin{equation}
       \frac{d\Phi_\nu}{dE_\nu} = 4.08\times 10^{19}\, f_{\rm DM} \,e^{\left(\frac{-1.539\times 10^{18}\, E_\nu^{3/2}}{\tau_{\rm DM} \, m_{\rm DM}^{3/2}} \right)}\frac{E_\nu^{1/2}}{\tau_{\rm DM} \, m_{\rm DM}^{5/2}}.       
       \label{eq:finflux2}
\end{equation}

Using this formalism, in the next section, we explore the possibility of explaining the KM3NeT and IceCube neutrino fluxes through the decay of superheavy dark matter produced via PBH evaporation.

\section{Interpretation of KM3NeT and IceCube Neutrino Observations}
\label{KM3 Interpretation}

In this section, we show that the framework of primordial black holes and neutrinos originating from the decay of dark matter particles emitted through PBH evaporation can be utilized to interpret the observed ultra-high-energy neutrino events, such as the recent KM3-230213A and IceCube detections. Using the expression for the neutrino flux given in Eq.~(\ref{eq:finflux2}), we calculate the flux of neutrinos originating in this framework, assuming a fixed superheavy dark matter mass of around ${\cal O}(100)$ PeV mass and various values of the lifetime of the DM $\tau_{\rm DM}$.  

As seen from Eq.~(\ref{f_j}), the phase space distribution of neutrinos depends on the number density of neutrinos at the time of PBH evaporation, \( n^{\mathrm{ev}}_i \). If neutrinos are produced through DM decay, their number density at evaporation will be limited by the DM number density at that time. The DM number density, \( n_{\mathrm{DM}} \), can be related to the comoving number density \( Y_{\mathrm{DM}} \) using Eq.~(\ref{niev}). Applying the relic abundance constraint \( \Omega_{\mathrm{DM}} \simeq 0.12 \), we obtain a corresponding constraint on \( Y_{\mathrm{DM}} \) from Eq.~(\ref{relic_ab_DM}). Substituting this into Eq.~(\ref{niev}), we compute the phase-space distribution and the resulting neutrino flux.

Interestingly, we find that the phase-space distribution, and consequently the resulting neutrino flux, does not explicitly depend on the PBH mass $M_{\rm BH_0}$ or on the parameter $\beta$. From Eq.~(\ref{f_j}), we note that only the parameters $n_i^{\rm ev}$ and $a_{\rm ev}^3$ depend on $M_{\rm BH_0}$ through Eqs.~(\ref{eq:Tev_rad}), (\ref{niev}), and (\ref{a_ev_rad}). However, since $n_{\rm DMev} \propto M_{\rm BH_0}^{-9/2}$ and $a_{\rm ev}^3 \propto M_{\rm BH_0}^{9/2}$, their product becomes independent of $M_{\rm BH_0}$, leading to a flux that does not explicitly depend on the PBH mass. 
\begin{figure*}[htbp]
    \centering
    \includegraphics[scale=0.6]{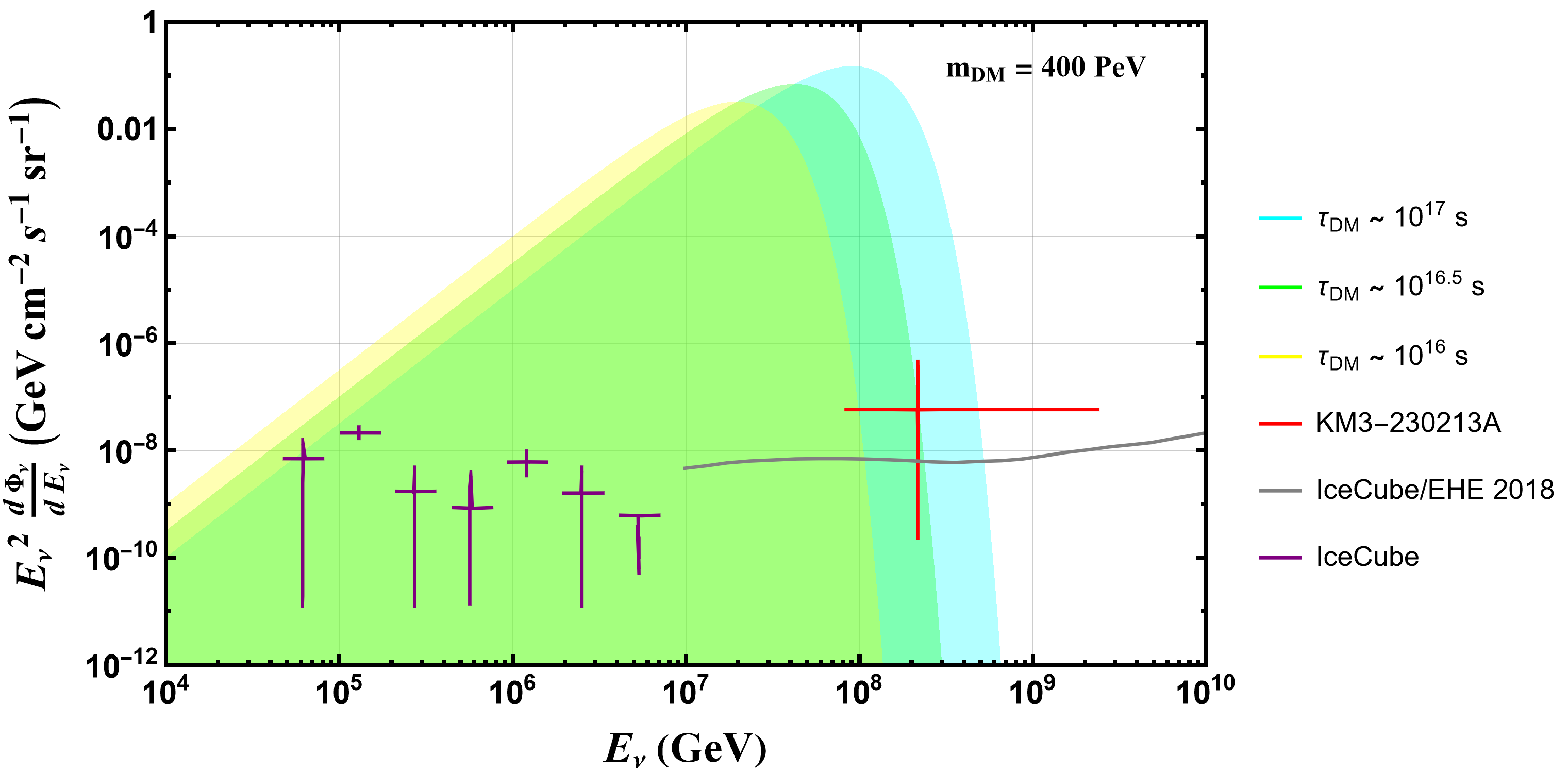}
     \caption{The energy squared differential neutrino flux per steradian with respect to the energy of the neutrinos. Figure shows the neutrino flux dependence on the DM lifetime $\tau_{\rm DM}$ at a constant mass of the dark matter $m_{\rm DM} =400\, {\rm PeV}$. The yellow, green and light blue shaded region corresponds to the flux for $\tau_{\rm DM}=10^{16}\,s$, $\tau_{\rm DM}=10^{16.5}\,s$ and $\tau_{\rm DM}=10^{17}\,s$ respectively. The violet and red data points correspond to observed high-energy neutrino events from or 7.5 years of IceCube data and recent KM3NET respectively. The grey line is the quasi-differential upperbound on the extremely high-energy neutrino flux observations from 9 years of IceCube observations.  }
     \label{Neutrino_flux_1}
\end{figure*} 
 
The constraints on $\beta$ and $M_{\rm BH_0}$ arise instead from fixing $Y_{\rm DM}$, as illustrated in the contour plot of Fig.~\ref{contourplot}. Figure~(\ref{Neutrino_flux_1}) shows the computed neutrino flux for different values of DM lifetime $\tau_{\rm DM}$, keeping DM mass $m_{\rm DM} = 400~{\rm PeV}$ fixed.  The theoretical results are compared with the experimental data, i.e., the red line corresponds to the recent high-energy KM3NeT observation~\cite{km3net2025observation}, grey line is the quasi-differential upperbound on the extremely high-energy (EHE) neutrino flux observations above $5\times 10^6$ GeV from the 9 years of IceCube observations~\cite{IceCube_EHE_2018} and the violet line corresponds to the high-energy all-sky neutrino observation of astrophysical origin denoted as IceCube High-energy Starting Event (HESE) for 7.5 years of data~\cite{IceCube_HESE}. As evident from the figure, for $\tau_{\rm DM} = 10^{16.5}\, s$ and $10^{17} \, s$, the framework can successfully account for both the IceCube and KM3NeT events. In particular, for $\tau_{\rm DM} = 10^{16.5}\,s$ and $E_\nu = 220~{\rm PeV}$, the predicted neutrino flux agrees remarkably well with the value reported by the KM3NeT Collaboration. 

\begin{figure*}[htbp]
    \centering
    \includegraphics[scale=0.6]{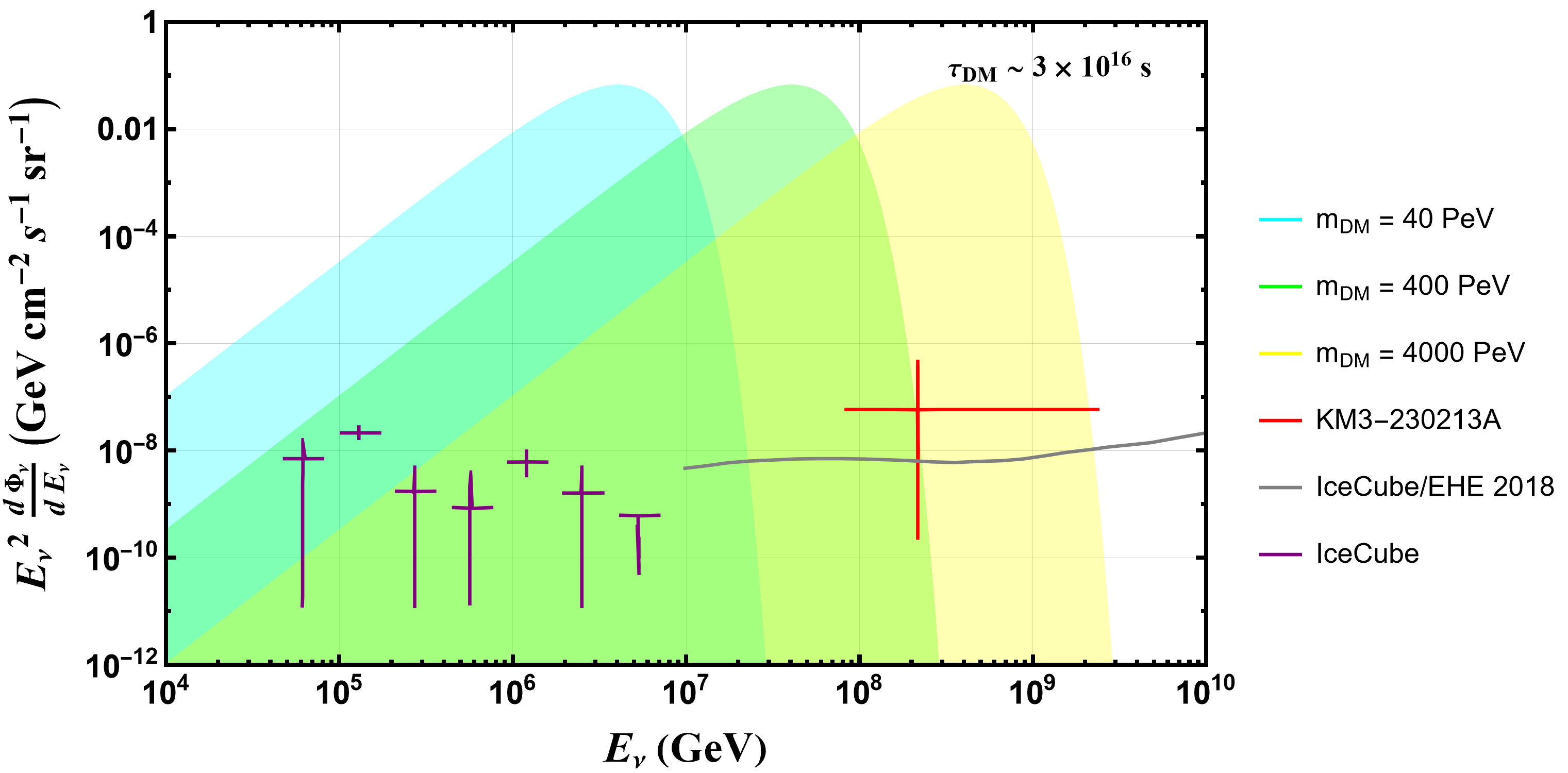}
\caption{The energy squared differential neutrino flux per steradian with respect to the energy of the neutrinos for a fixed DM lifetime $\tau_{\rm DM} = 3 \times 10^{16}\,s$. The yellow , green and light blue shaded region corresponds to the flux for $m_{\rm DM} = 4000~{\rm PeV}, 400~{\rm PeV}$ and $40~{\rm PeV}$ respectively. The violet and red data points correspond to observed high-energy neutrino events from or 7.5 years of IceCube data and recent KM3NET respectively. The grey line is the quasi-differential upper bound on the extremely high-energy neutrino flux observations from 9 years of IceCube observations.}
    \label{Neutrino_flux_2}
    \end{figure*}

\begin{figure*}[htbp]
    \centering
    \includegraphics[scale=0.6]{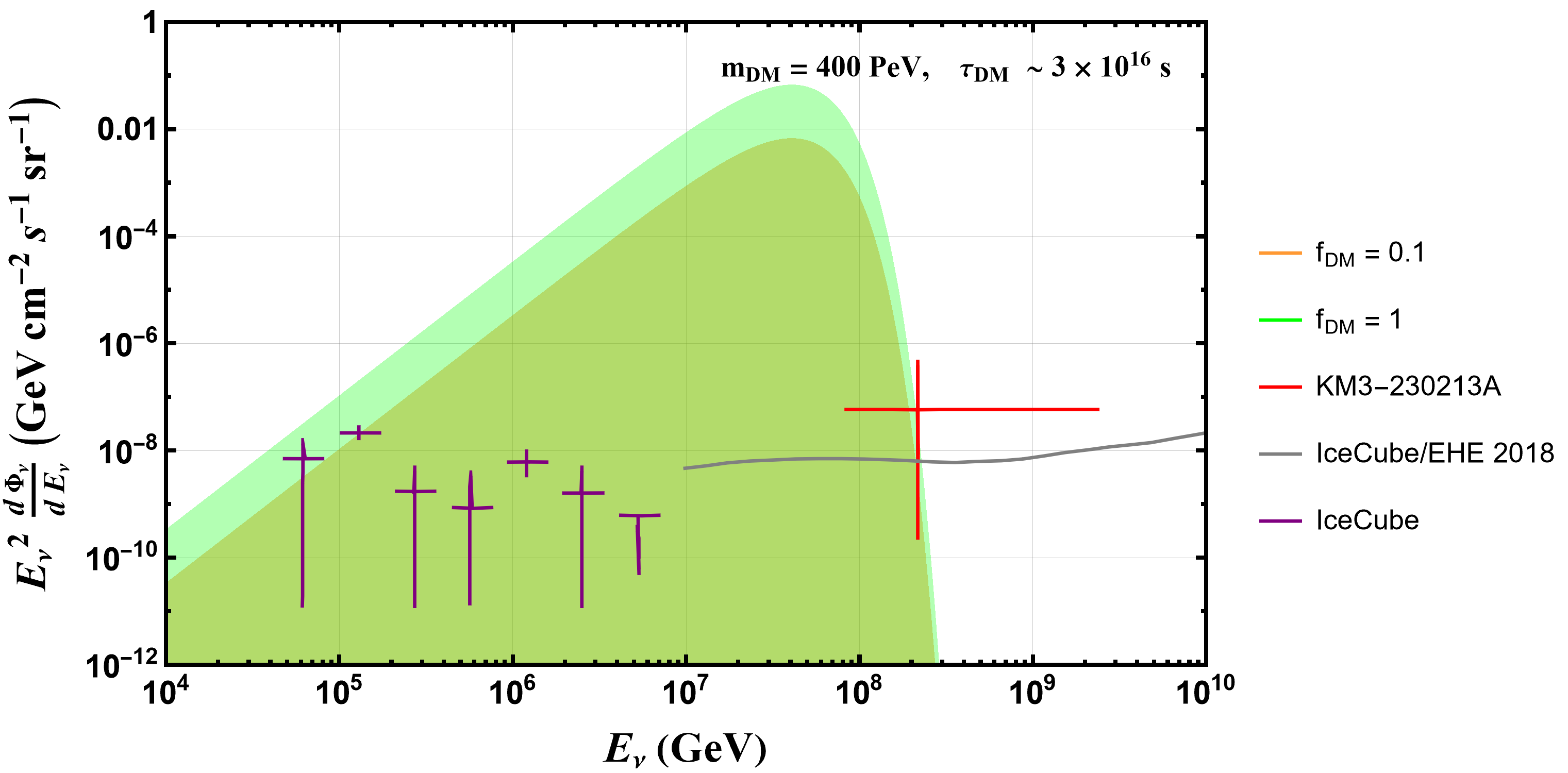}
    \caption{The neutrino flux as a function of PBH constituting the fraction of total dark matter $f_{\rm DM}$ at fixed mass of the dark matter $m_{\rm DM}=400\, {\rm PeV}$ and lifetime of the DM $\tau_{\rm DM} = 3\times 10^{16}\,s$.}
    \label{Neutrino_flux_fDM}
\end{figure*}

Furthermore, as shown in Fig.~\ref{contourplot}, fixing the dark matter mass at $m_{\rm DM} = 400~{\rm PeV}$ yields an allowed range for the PBH mass $M_{\rm BH_0}$ between $\sim 10~{\rm g}$ and $ 10^8~{\rm g}$, and for the parameter $\beta$ between $\sim 10^{-24}$ and $ 10^{-14}$. Hence, we conclude that within a broad range of PBH masses and values of $\beta$, the proposed framework can consistently explain the origin of both the IceCube and KM3NeT high-energy neutrino events.
     In Fig.~\ref{Neutrino_flux_2}, we present the neutrino flux for different values of the dark matter mass while keeping the lifetime of DM fixed at $\tau_{\rm DM} = 3 \times 10^{16}\,s$. The results indicate that increasing the dark matter mass shifts the flux toward higher energies. 
In Fig.~\ref{Neutrino_flux_fDM}, we show the neutrino flux for different fractions of the DM relic density, keeping both the DM mass and the DM lifetime fixed. The results demonstrate that the decrease in the DM density fraction leads to a corresponding suppression in the neutrino flux.
\paragraph{Constraints on $\beta$:}
From the previous discussion, Figs.~\ref{Neutrino_flux_1},~\ref{Neutrino_flux_2}, and~\ref{Neutrino_flux_fDM} indicate that the observed high-energy neutrinos can be consistently interpreted as originating from the evaporation of primordial black holes into superheavy DM  particles with masses in the range $10^{7} \mathrm{GeV} \leq m_{\rm DM} \leq 10^9 \mathrm{GeV}$. From Fig.~\ref{contourplot}, corresponding to this mass range, the allowed parameter space for the initial PBH mass $M_{\rm BH_0}$ and PBH abundance $\beta$ can be determined.

A variety of astrophysical and cosmological observations impose upper limits on the initial PBH abundance parameter $\beta$. A comprehensive summary of various astrophysical and cosmological constraints on the PBH mass range $\mathrm{M_{BH}}_0 \in [1~\mathrm{g}, 10^{50}~\mathrm{g}]$ is provided in Fig.~9 of Ref.~\cite{carr2010_constraint_PBH}.  While these constraints are relatively weaker for low-mass PBHs, stringent bounds arise from CMB spectral distortions and anisotropies, which restrict the allowed values of $\beta$. The corresponding upper limit is given by~\cite{carr2010_constraint_PBH} 
\begin{equation}
    \beta\,'<10^9\left(\frac{M}{M_{P}}\right)^{-1}\approx10^{-5}\left(\frac{M}{10^9\, \rm g}\right)^{-1},\,\,\,(M<10^9\,\rm g).
\end{equation}
Here, the $\beta\,'$ is a new parameter depending on $\beta$ as,
\begin{equation}
    \beta\,'(M) \equiv \gamma^{1/2}\,\left(\frac{\rm g_{*i}}{106.75}\right)^{-1/4}\, \beta(M),
\end{equation}
where $g_{*i}$ is the number of relativistic degrees of freedom at the time of PBH formation. For  $g_{*i} \sim g_{*\rm SM}$,   $\beta\,'(M) \approx \beta(M)$. One can see that, for PBH of mass $<10^4\,\rm g$, the value of $\beta\,$ is greater than $1$, i.e., the energy density of PBH is greater than the universe. This would result in the PBH/early matter-dominated universe. Since we are only focusing on the radiation-dominated universe for this paper, PBH mass in the range from $10^4 \,\rm g <  M_{BH_0}< 10^{9}\, g$ can be constrained. For these range of $\rm m_{DM}$ and $\rm M_{BH_0}$, the $\beta$ ranges from $10^{-20} < \beta < 10^{-12}$.

Further constraints arise from the emission of unstable particles, such as gravitinos~\cite{khlopov_1,Khlopov_3} or neutralinos, during PBH evaporation. The decay of these particles into lighter species can significantly modify the abundances of light elements predicted by BBN, thereby placing additional and stringent bounds on $\beta'$~\cite{carr2010_constraint_PBH}.
\begin{eqnarray}  
&&  {\hskip -0.2in} \beta\,'(M) \lesssim 5\times 10^{-19}\,\left(\frac{M}{10^9\, \rm g}\right)^{-1/2}\left(\frac{Y}{10^{-14}}\right)\left(\frac{x_\phi}{0.006}\right)^{-1},\notag \nonumber\\
&&  {\hskip -0.2in} \rm for\,\,(M<10^9\,\rm g). 
\end{eqnarray}
Here, $Y$ denotes the upper limit on the particle yield, and $x_\phi$ represents the fraction of the PBH luminosity emitted into quasi-stable massive particles. 

In our results, as the required neutrino flux can be obtained for a broad range of PBH masses, $M_{\rm BH_0} \in [10~\mathrm{g}, 10^{8}~\mathrm{g}]$, and PBH abundances, $\beta \in [10^{-24}, 10^{-14}]$, by considering a superheavy DM mass of $m_{\rm DM} \sim 400~\mathrm{PeV}$, the parameter space remains viable even after accounting for the aforementioned astrophysical and cosmological constraints. Consequently, the proposed framework can consistently explain the high-energy neutrino events observed by both IceCube and KM3NeT.
\section{Summary and Future Directions}
\label{Conclusions}

We have explored a unified framework in which the recently observed ultra-high-energy neutrino event KM3-230213A, together with similar IceCube detections, can originate from the late-time decay of metastable superheavy DM particles produced through PBH evaporation in the early Universe. Within this scenario, PBHs formed from primordial density perturbations evaporate via Hawking radiation, emitting both SM particles and long-lived superheavy DM candidates. The subsequent decay of these DM particles at late cosmological epochs gives rise to a neutrino flux observable at present times.

Using the observed DM relic density as a constraint, we determined the viable regions of the PBH abundance parameter $\beta$ and the initial mass $M_{\rm BH_0}$ that yield the correct DM production. For $10\,{\rm g}\!\lesssim\!M_{\rm BH_0}\!\lesssim\!10^{8}\,{\rm g}$ and $10^{-24}\!\lesssim\!\beta\!\lesssim\!10^{-14}$, the decay of DM with masses $m_{\rm DM}\!\sim\!10^{7}$--$10^9\,{\rm GeV}$ can reproduce the PeV-EeV neutrino flux consistent with current observations. The proposed mechanism also naturally avoids stringent multimessenger constraints and remains compatible with cosmological and astrophysical bounds.

Overall, this study establishes PBH evaporation as a compelling mechanism for generating superheavy DM and a plausible explanation for the observed ultra–high–energy neutrino events. Future observations from KM3NeT and IceCube-Gen2 will be instrumental in testing the predicted neutrino flux and assessing the viability of this PBH-induced dark matter decay framework. While the decay of PBH induced massive DM avoid the standard multimessenger signals, it should be noted that indirect production of gamma rays can happen due to electroweak bremsstrahlung by emission of W and Z bosons, aside from photons. From the decay of these W and Z gauge bosons, secondary annihilation particles like electrons and positrons, protons and anti-proton as well as neutrinos and other charged leptons and hadrons may be produced~\cite{ElectroWeakBS1,electroweakBS2}. The resulting gamma rays may therefore be observed by various observatories like LHAASO~\cite{LHAASO:2023rpg}, Tibet $\rm AS\gamma$~\cite{TibetASgamma}, HAWC~\cite{HAWC:2019tcx}, and Fermi-LAT~\cite{Fermi-LAT:2009ihh}. In addition, complementary probes from intermediate cosmic epochs such as constraints on energy injection derived from the 21–cm signal can further test this scenario~\cite{Sun:2025ksr}, offering a unique opportunity to bridge early universe cosmology with present-day high-energy neutrino astrophysics.
\\
\section{Acknowledgment}
\label{Acknowledgment}
PS and MD acknowledge the organizers of the Vikram Discussion on Neutrino Astrophysics (February 2025), held at the Physical Research Laboratory (PRL), Ahmedabad, India, for organizing insightful discussions and talks on high-energy neutrinos and KM3NeT, which were helpful in the completion of this paper. PS also thanks Prof. Srubabati Goswami for her valuable comments and suggestions.

\appendix 
\section{Relation between $a_d$ and $\tau_{\rm DM}$}
 \label{appendix:1}
 We can write the Hubble parameter in terms of scale factor in case of matter dominated era as,
$$H = H_0 \sqrt{\Omega_{m,0}\,a^{-3}}$$
where $\Omega_{m,0}$ and $H_0$ are the relic abundance and Hubble parameter today given as, $\Omega_{m,0} = 0.315$ and $H_0 = 67.4\, {\rm km}\, {\rm sec}^{-1}\, {\rm Mpc}^{-1} = 1.44\times 10^{-42}\, {\rm GeV}$~\cite{Planck_sat_DM}. Substituting above constants, we get the Hubble parameter as
$$\frac{H}{\rm GeV} = 8.066\times10^{-43}\sqrt{\frac{1}{a_d^3}}. $$
Now, equating the Hubble expansion rate ${\rm H}$ with the decay rate of DM, i.e., $H \sim \Gamma_D = \hbar/\tau_{\rm DM}$, we get, 
$$\tau_{\rm DM} = \frac{\hbar}{\Gamma_D},$$
where $\tau_{\rm DM}$ is the lifetime of DM and the $\hbar$ is reduced Planck constant given by $\hbar = 6.583\times 10^{-25} \, {\rm GeV}\,{\rm sec}$. Using this, we get
$$\tau_{\rm DM} = 2.581\times 10^{16}\, \left(\frac{a_d}{0.1}\right)^{3/2}\,{\rm sec}.$$
From this, the energy flux of the ultra-high-energy neutrino given in Eq.~(\ref{eq:finflux}) can be expressed in terms of the dark matter lifetime as follows:
$$  \frac{d\Phi_\nu}{dE_\nu} = 4.08\times 10^{19}\, f_{\rm DM} \,e^{\left(\frac{-1.539\times 10^{18}\, E_\nu^{3/2}}{\tau_{\rm DM} \, m_{\rm DM}^{3/2}} \right)}\frac{E_\nu^{1/2}}{\tau_{\rm DM} \, m_{\rm DM}^{5/2}}.
       $$
 
\bibliographystyle{unsrt}

\end{document}